\documentclass[aps,prb,floatfix,twocolumn,showpacs,preprintnumbers,superscriptaddress,amsmath,amssymb]{revtex4-1}

\usepackage{graphicx}  
\usepackage{dcolumn}   
\usepackage{bm}        
\usepackage{stmaryrd}  
\usepackage{multirow}  
\usepackage{color}     
\usepackage{amsmath,amssymb}
\usepackage{enumerate}
\usepackage{float}
\usepackage{mathtools}

\newcommand{\der}[3]{\frac{{\rm d}^{#1} #2}{{\rm d} #3^{#1}}}     
\newcommand{\pder}[2]{\frac{\partial #1}{\partial #2}}     
\newcommand{\vder}[2]{\frac{\delta #1}{\delta #2}}     

\def   \dd      {{\rm d}}

\def   \ez      {\hat{\bm e}_z}
\def   \Sext    {S^{\rm ext}}

\def   \det      {\widetilde{\bigl[ \Delta t \bigr]}}
\def   \dett     {\widetilde{\biggl[ \frac{\Delta t}{\tau} \biggr]}}

\def   \zs       {z_{\rm s}}
\def   \ts       {t_{\rm s}}
\def   \avn      {\bar{n}}
\def   \emax     {\epsilon_{\rm max}}

\def   \Nbar    {\bar{N}}

\def \js {{j_{\rm s}}}

\def \bM {{\bm M}}
\def \Ms {{M_{\rm s}}}

\def \gg   {{\gamma_{\rm g}}}
\def \btau {{\bm \tau}}
\def \ex   {{\hat{\bm e}_x}}

\def \ez   {{\hat{\bm e}_z}}

\begin{document}

\preprint{APS/123-QED}

\title{Transport theory for femtosecond laser-induced spin-transfer torques}

\author{Pavel Bal\'a\v{z}} 
\affiliation{Charles University, Faculty of Mathematics and Physics, Department of Condensed Matter Physics, Ke Karlovu 5, CZ-121 16 Prague, Czech Republic}
\author{Martin \v{Z}onda} 
\affiliation{Charles University, Faculty of Mathematics and Physics, Department of Condensed Matter Physics, Ke Karlovu 5, CZ-121 16 Prague, Czech Republic}
\author{Karel Carva} 
\affiliation{Charles University, Faculty of Mathematics and Physics, Department of Condensed Matter Physics, Ke Karlovu 5, CZ-121 16 Prague, Czech Republic}
\author{Pablo Maldonado} 
\affiliation{Department of Physics and Astronomy, Uppsala University, Box 516, SE-75120 Uppsala, Sweden}
\author{Peter M. Oppeneer} 
\affiliation{Department of Physics and Astronomy, Uppsala University, Box 516, SE-75120 Uppsala, Sweden}

\date{\today}

\begin{abstract}
  Ultrafast demagnetization of magnetic layers pumped by a femtosecond laser pulse is  accompanied by a nonthermal spin-polarized current of hot electrons. 
  These spin currents are studied here theoretically in a spin valve with noncollinear magnetizations. 
  To this end, we introduce an extended model of superdiffusive spin transport that enables to treat noncollinear magnetic configurations, 
  and apply it to the perpendicular spin valve geometry.
  We show how spin-transfer torques arise due to this mechanism and calculate their action on the magnetization present,
  as well as how the latter depends on the thicknesses of the layers and other transport parameters.
  We demonstrate that there exists a certain optimum thickness of the
  out-of-plane magnetized spin-current polarizer such that the torque acting on the second magnetic layer is maximal.
  Moreover, we study the magnetization dynamics excited by the superdiffusive 
  spin-transfer torque due to the flow of hot electrons employing the Landau-Lifshitz-Gilbert equation.
  Thereby we show that a femtosecond laser pulse applied to one magnetic layer can excite
  small-angle precessions of the magnetization in the second magnetic layer. 
  We compare our calculations with recent experimental results.
\end{abstract}


\maketitle

\section{Introduction}

The ultrafast demagnetization of a Ni thin film induced by a femtosecond laser pulse \cite{Beaurepaire1996:PRL} marked a real breakthrough
in the research on dynamic manipulation of magnetic moments. Subsequently,  a number of exciting research studies have been conducted in this area.
Along one of the current lines of investigations, researchers try to develop viable schemes to manipulate magnetic moments with ultrashort laser pulses.
Especially, an efficient method for all-optical switching (AOS) of magnetization became the desired goal of many research projects, 
which has been successfully achieved thus far for several materials.~\cite{Stanciu2007,Mangin2014,Lambert2014,Stupakiewicz2017_YIG_Co}
Another main line of research has concentrated on the foundational explanation of laser-induced effects in magnetic materials, which has become a recurrent theme of 
many scientific papers.~\cite{Zhang2000,Malinowski2008,Koopmans2010,Battiato2010:PRL,Carva2011PRL,Carva2011_NPhys,Illg2013,Haag2014,Berritta2016}
It is possible that there is not just a single physical mechanism responsible for the laser-induced demagnetization and that the contributions of the various mechanisms would depend on which materials are involved. \cite{Turgut2013,Schellekens2014PRB,Bergeard2016,Turgut2016}
The theoretical description of Battiato, Carva, and Oppeneer~\cite{Battiato2010:PRL,Battiato2012:PRB} proposes that the loss of magnetic moment in a magnetic layer after 
the laser irradiation is caused by the transport of the laser-excited electrons into adjacent metallic layers or substrate.
The asymmetry of the transport properties with respect to the electron spin orientation in magnetic materials then leads effectively to a superdiffusive
spin current (SC). 
This nonequilibrium SC not only reduces the magnetization of the pumped layer, but has also been shown experimentally to affect adjacent magnetic layers~\cite{Rudolf2012,Eschenlohr2017:JPhys} 
or to generate THz spin-current pulses. \cite{Kampfrath2013}
Notably, spin currents represent a key component of spintronics, and control over femtosecond SCs is highly desirable for potential
spintronic applications operating at THz frequencies.

Spin currents can be generated by charge currents in multilayers by combining sufficiently thin magnetic and nonmagnetic metallic films. 
This can give rise to the giant magnetoresistance (GMR) effect,~\cite{r_88_Baibi_Fert_GMR,r_89_bgs}
which has been explained by means of the spin-dependent diffusive transport model,~\cite{Camley1989:PRL,Barnas1990:PRB}
and later been successfully calculated by \textit{ab initio} methods. \cite{r_03_pw,r_06_ctkb_vertex}
Here the resistance depends significantly on the magnetic configuration of the system and it can thus be used to read information stored in it.
There is also a complementary effect; the spin angular momentum transported between the magnetic films can induce magnetic excitations
in the magnetic layers, which can lead to a change of the magnetic configuration of the spin valve.~\cite{Slonczewski1996,Berger1996}
In this process the current of flowing electrons is polarized in the first magnetic layer
and then it is transported through the nonmagnetic spacer.
When it reaches the second magnetic layer, the component of the spin current, which is collinear to the
local magnetization enters the magnetic layer.
However, the noncollinear component---the transverse spin current---is absorbed at the interface between the nonmagnetic and magnetic layers and
gives rise to the so called spin-transfer torque (STT).
Consequently, the local magnetization of the second layer changes its direction.
The typical length scale at which the transverse spin current is absorbed in the magnetic layer is
at most few nanometers.~\cite{r_02_StilesAnatomy,r_02_Slon_CurrTor,r_12_Ghosh_PenetrDepth_SpinCurr,r_16_Balaz_Zwier_TransSpinPene_SHM,Lalieu2017:PRB}
Clearly, the STT appears only in the case when the magnetizations of the two adjacent magnetic layers are noncollinear.
The effect of the STT can be described in a way analogous to the Valet-Fert~\cite{r_93_vf}
model using the diffusive spin transport description taking into account bulk resistivities and interface conductances.~\cite{Barnas2005:PRB}
Both these effects can be utilized in magnetic random access memories (MRAM) and together with other phenomena, mostly based on the spin-orbit interaction, have greatly stimulated the field of spintronics.\cite{r_04_Zutic_RevSpintronics,Sinova2015}

In contrast to magnetic spin valves operated by dc currents, the laser pulses applied to a magnetic layer
provide ultrashort SC pulses on the timescale of the demagnetization -- femtoseconds. Moreover, the energy distribution of the flowing electrons is different.
While in the standard spin valves the spin current flows solely at the Fermi level,
in the case of laser-excited SC the energy distribution is more complex. Here the nonthermal electrons are located at energies of about one eV above the Fermi level after being first excited from the $d$ band to the $sp$ band. 
Since the electron mobility in the $sp$ band is much higher than in the $d$ band, the electrons quickly 
move away from the laser spot. \cite{Battiato2010:PRL,Battiato2012:PRB}
We note that SC generation due to a different mechanism, electron-magnon scattering and spin-pumping without the need for far-from-equilibrium hot electrons  has also been suggested. \cite{Choi2014:NatComm} In contrast to that, a recent study points out the importance of the scattering of hot electrons at the interfaces for the form of the spin current induced by ultrafast demagnetization. \cite{Alekhin2017_fsSC_pulse}
Despite the different thermal and nonthermal mechanisms and corresponding timescales, 
it has been shown by recent experimental studies~\cite{Schellekens2014:NatComm,Choi2014:NatComm,Lalieu2017:PRB} 
that in such laser-excited magnetic spin valve one observes an ultrafast STT acting on the second magnetic layer.

Notably, laser excitation is always accompanied by the appearance of a temperature gradient along the heated ferromagnetic layer, when the excited hot electrons have thermalized, a process that may take a few hundred femtoseconds.
The electrons in the thermal gradient are then transported in separated spin channels with different spin resistivities,
which also leads to the emergence of a SC.
Therefore, the spin-dependent Seebeck effect~\cite{Choi2015:NatMater} comes under discussion. 
However, this effect becomes dominant only on a significantly slower timescale than the effect of the ultrafast demagnetization 
and hence it is possible to distinguish these two contributions.\cite{Choi2015:NatMater,Carva2014_NPhys}

The goal of this paper is twofold. 
First, we shall extend the theoretical description of the superdiffusive transport.\cite{Battiato2010:PRL,Battiato2012:PRB}
The original model assumes that hot electrons are transported throughout the layered structure in two spin channels.
This assumption, however, limits the numerical simulations strictly to collinear relative configurations of magnetizations in the magnetic layers;
i.e., all the magnetizations must be aligned along the same axis and the magnetizations of the neighboring magnetic layers
can only be either parallel or antiparallel. The orientation of local magnetization is reflected in the transport properties
of the two spin channels.
Evidently, transport in a collinear configuration is sufficient to describe the effects related to ultrafast demagnetization, however, the effect of spin-transfer torques is strictly linked to noncollinear magnetic configurations.
Therefore, we extend here the superdiffusion model to be able to account for noncollinear magnetic configurations, as shall be explained in Sec.~\ref{Sec:Model}.
Shortly, in our extended model, we assume a magnetic multilayer consisting of magnetic and nonmagnetic layers, where the magnetizations of the magnetic layers
can be either in-plane or perpendicular to the plane. 
In addition, we shall assume that the magnetization dynamics is rather weak and does not substantially influence the
transport properties of the electrons. 
Thus, we can describe the spin transport assuming the presence of electrons with two possible fixed quantization axes.
In a nonmagnetic layer, all spin orientations are equivalent. 
In a magnetic layer, however, electrons with spins aligned along the local magnetization axis contribute to the longitudinal spin current,
whereas the perpendicularly oriented spins are the sources of a transverse spin current that plays an important role.
It is the latter one that is rapidly absorbed in the magnetic layer (with perpendicular magnetization axis) and gives rise to the STT.

Second, we focus on a typical spin valve composed of two ferromagnetic films separated by a nonmagnetic one.
One of the magnetic layers, which will be directly irradiated by the laser pulse, has perpendicular-to-plane magnetization, 
while the second one is in-plane.
Using our extended model of superdiffusive spin-dependent electron transport we quantify the STT
acting on the in-plane magnetization as well as its dependence on the transport parameters and spin valve geometry.
Using the Landau-Lifshitz-Gilbert equation (LLG) we examine the laser-induced magnetization dynamics of the in-plane magnetization.
We show that a femtosecond laser pulse applied on the layer with perpendicular magnetization can excite precessions of the in-plane magnetization
persisting up to few nanoseconds triggered by the transverse superdiffusive SC.

This paper is organized as follows.
In Section~\ref{Sec:Model} we extend the superdiffusive spin-dependent transport model for noncollinear configurations
and define the spin-transfer torque acting on the magnetizations.
In Sec.~\ref{Sec:Results} we describe our results for the transverse spin current and the spin dynamics in a noncollinear spin valve geometry.
Finally, we summarize the most important findings in Sec.~\ref{Sec:Conclusions}.

\section{Model}
\label{Sec:Model}

\begin{figure}[t!]
  \centering
  \includegraphics[width=.9\columnwidth]{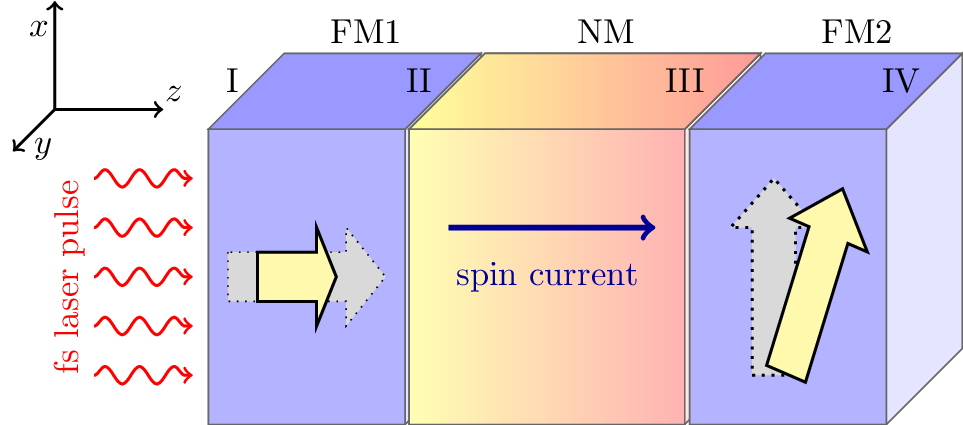}
  \caption{(Color online) Sketch of the magnetic spin valve consisting of two magnetic layers, FM1 and FM2, separated by a nonmagnetic one, NM.
           The dotted (gray) arrows present the initial magnetic configuration before the illumination by the laser pulse.
           After the laser pulse is applied, the magnetization in FM1 decreases and gives rise to the spin current in NM.
           Consequently, due to the spin-transfer torque the magnetization in FM2 is tilted from its equilibrium position.
           This nonequilibrium situation is depicted by the solid (yellow) arrows.
           The roman numbers enumerate the interfaces which are located at positions $z_{\rm I} = 0$, $z_{\rm II}$, $z_{\rm III}$, and $z_{\rm IV}$.}
  \label{Fig:scheme}
\end{figure}
We assume a metallic spin valve consisting of two ferromagnetic layers, FM1 and FM2, separated by a nonmagnetic one, NM, as shown in Fig.~\ref{Fig:scheme}.
The first magnetic layer (FM1) has magnetization perpendicular to the plane of the layers. 
Such a magnetic configuration can be experimentally achieved using a composite layer composed of repeated sequences of ultrathin magnetic layers such as Co/Ni or Co/Pt 
where the thicknesses of the sublayers are as large as a few \AA{}ngstr\"oms.
In such systems the perpendicular magnetic anisotropy results in an out-of-plane orientation of magnetic moments.~\cite{Broeder1992:TransMagn,Gottwald2012:PRB}
In this work we shall treat this magnetic layer as homogeneous. This simplification allows us to model the electronic transport
inside the layer by means of the superdiffusive model, as we shall describe below.
The second magnetic layer (FM2) has its magnetization in the layer's plane, which is in equilibrium aligned along the magnetization easy axis.

\subsection{Electronic transport in noncollinear systems}

The model of superdiffusive spin-dependent transport has been developed to explain ultrafast demagnetization in metallic heterostructures.~\cite{Battiato2010:PRL,Rudolf2012}
The model assumes that the laser pulse excites electrons from the quasilocalized $d$-band to the $sp$-band above the Fermi level.
Since the mobility of the $sp$-electrons is higher they start to move through the heterostructure.
One of the main assumptions of the model is that each energy level is spin degenerate and
the electrons move through the layered structure in two spin channels.
Importantly, in the framework of this two channel model, presented in previous works,~\cite{Battiato2010:PRL,Battiato2012:PRB} 
one can model only spin transport in \textit{collinear} magnetic configurations. 
However, to generate transverse spin-current components (with respect to some of the local magnetizations),
resulting in spin-transfer torque, a noncollinear magnetic configuration is required.
To this end, in this paper we extend the superdiffusive transport model to be able to study noncollinear magnetic configurations.

Initially, for the sake of simplicity we make several assumptions.
First, in our simplified model, we assume just a static magnetic configuration. 
When a spin-transfer torque is acting on the magnetization it tends to change its direction.
In case of spin valves or tunnel junctions, the spin torque generated by a continuously flowing current can
induce large angle magnetization precessions or even switch the magnetization direction.
In case of magnetic multilayer devices operated by laser pulses the situation observed in experiments is different.
The major change of magnetization due to a laser pulse is governed by the longitudinal spin relaxation, 
which changes only the magnetization length keeping its direction unchanged.
Consequently, if a transverse spin current builds up at the interface between the nonmagnet and ferromagnet, it exerts a torque on the 
magnetic moments inducing the spin dynamics.
However, in comparison to electric field operated devices, the laser-pulse induced magnetization dynamics is a feeble effect
occurring as small-angle precessions of magnetic moments around their initial directions.
Therefore we assume that the magnetization changes during the spin dynamics do not particularly influence the 
spin-dependent electronic transport.
As a result, we can calculate first the time evolution of spin currents and spin-transfer torques acting on the magnetic moments of the multilayer and
then we separately model the laser-induced spin dynamics.

We also assume that spin-transfer torque due to hot electrons acts as an interfacial effect.
Let us first discuss here possible relevant theories for spin-transfer torque, and the range of validity of our approach.  
Importantly, spin transport in systems with noncollinear magnetizations rest on inherently quantum phenomena related to the fact 
that electrons with spin orientation different from the local quantization axis are not eigenstates of itinerant electrons.
This effect can be calculated in terms of spinor wave functions connected at interfaces via scattering matrices~\cite{r_00_Wain_Torque}.
It has been also described by the magnetoelectronic circuit theory~\cite{r_00_bnb,r_06_aBraBK_NonColMagElectr}, 
which employs charge currents and three-component spin currents between nodes of the system. 
These are related to the general $2 \times 2$ current in spin space
$\hat{\bm \jmath} = j_0 \hat{\bm 1} + {\bm j} \cdot {\bm \sigma}$, 
where $j_0$ is the charge current, ${\bm j} = (j_x,j_y,j_z)$ is the spin current vector, and ${\bm \sigma}$ is a vector of Pauli matrices.
This formalism is sufficient to describe all crucial effects in noncollinear magnetic systems.
These quantum mechanical methods cannot be directly combined with the semiclassical superdiffusive transport model.
However, we utilize their findings in our model. 
In the non-magnetic part the equations for spin currents have trivial structure in spin space as long as spin flips are neglected, 
and multiple equivalent superdiffusive equations applied independently to a sufficient number of spin channels 
describe correctly propagation of spin currents there. 
In the FM part spin currents parallel to the FM orientation can be described employing two inequivalent equations (two channel model). 
On the other hand, the transverse spin current undergoes rapid dephasing, which leads to violation of spin current conservation. Furthermore, 
theory expects the transverse spin current to decay over a few atomic distances from the interface \cite{r_02_Slon_CurrTor}. A recent experimental study shows that the laser-induced transverse spin current is absorbed in ferromagnets on a distance
as short as $\sim 2\, {\rm nm}$,~\cite{Lalieu2017:PRB}. 
An experiment employing spin pumping also found the transverse spin current penetration depth to be below 2nm~\cite{r_12_Ghosh_PenetrDepth_SpinCurr}.
The only contribution to equations for spin currents that connects noncollinear spin currents, and requires extra source/sink term for transverse spin current in superdiffusive equations, 
can thus be localized and solved just at the interfaces between NM and FM parts. 
Notably, the longitudinal spin current and the total charge current remain conserved there.
More specifically, we assume the complete absorption of transverse spin current within the thicknesses of FM layers considered here (larger than 2nm). 
This can be also formulated as the limit of saturated mixing conductance, with its real part value close to the Sharvin conductance.
In previous works~\cite{r_07_ct_mixcond,r_05_ztk} it has been shown by first principles calculations 
that for realistic (not ideal) materials this limit is often achieved within few monolayer thickness.
The spin torque acting on the FM2 (FM1) magnetization is then proportional to the transverse spin current components at the NM/FM2 (FM1/NM).

We distinguish between the longitudinal and transverse spin components by assigning two separate spin channels to each of them, hence four channels in total are used.
This enables us to model spin-dependent transport in a perpendicular magnetic configuration of a spin valve as shown in Fig.~\ref{Fig:scheme}.
Two channels (with spins aligned along the FM1 magnetization) are assigned for electrons with spins aligned along  the FM1 magnetization axis, while
the remaining two channels transport electrons with spins parallel and antiparallel to the FM2 magnetization axis.
In the nonmagnetic layer, without any natural quantization axis, all the transport channels are equivalent featuring the same 
electron velocities and lifetimes at the same energy levels. 
However, when the electrons enter a magnetic layer with magnetization axis that does not agree with the direction of their spin
the electron spin rapidly relaxes, and thus only two channels are important for the description of electronic propagation there.
In the superdiffusive transport model, the fast relaxation of the transverse spins can be achieved by setting short lifetimes and
small velocities at all energy levels for hot electrons in the transverse spin channels.

\subsection{Mathematical formulation}

The transport of hot excited electrons in each layer (magnetic or nonmagnetic) can be described by a number of distinct energy levels with energies $\epsilon_i$, where $i = 1, 2, \dots, n$.
Furthermore, each energy channel with energy $\epsilon$ is divided into four distinct spin channels labeled as $\sigma \in \{\uparrow, \downarrow, \leftarrow, \rightarrow\}$.
The spin transport in each channel is labeled in the same way.
In each spin channel, electrons have lifetimes, $\tau_\sigma = \tau_\sigma(\epsilon,z)$, and velocities, 
$v_\sigma = v_\sigma(\epsilon,z)$,~\cite{Battiato2014:JAP} where $z$ is the position of the electron.
The position dependence of the four quantities suggests their difference in various materials.
These parameters are accessible by ab initio calculations~\cite{r_06_Zhuko_IMFP_GW_FeNi}
which can be further corrected taking into account the electron-phonon interaction.~\cite{Battiato2012:PRB}

Following Ref.~\onlinecite{Battiato2014:JAP}, we write the superdiffusive transport equation in each spin channel inside the layers as
\begin{equation}
  \begin{split}
  \pder{}{t}n_\sigma + \frac{n_\sigma}{\tau_\sigma} =
  &-\pder{}{z}\, \hat{\phi} \left(\frac{p_\sigma\, n_\sigma}{\tau_\sigma} + \Sext_\sigma\right) \\
  &+ \frac{p_\sigma\, n_\sigma}{\tau_\sigma} + \Sext_\sigma\,,
  \end{split}
\label{Eq:sde}
\end{equation}
where $n_\sigma = n_\sigma(z,t,\epsilon)$ is the particle density in spin-channel $\sigma$, 
$\Sext_\sigma = \Sext_\sigma(z,t,\epsilon)$ is the external source of particles for channel $\sigma$,
$p_\sigma = p_\sigma(z,\epsilon)$ is the probability that particle with spin $\sigma$ is not stopped after scattering.
For the sake of clarity, in the equations below, we shall omit the energy-dependence of the physical quantities,
unless it is necessary for the explanation.
Moreover, the spin-dependent operator $\hat{\phi}_\sigma$ acting on a general spin-dependent source term $S_\sigma(z,t)$ is defined as
\begin{equation}
  \begin{split}
    &\hat{\phi}_\sigma\, S_\sigma(z,t) = \\
    &\int_{-\infty}^{+\infty} \dd z_0 \int_{-\infty}^{t} \dd t_0\; S_\sigma(z_0,t_0)\, \phi_\sigma(z,t|z_0,t_0)\,,
  \end{split}
\label{Eq:phi_op}
\end{equation}
where the spin-dependent flux kernel reads
\begin{equation}
  \begin{split}
  &\phi_\sigma(z,t|z_0,t_0) =\; \\
  &\frac{\det_\sigma}{2 (t - t_0)^2}\; 
   \exp \left\{ -(t-t_0) \dett_\sigma \biggm/ \det_\sigma \right\}\, \times \\
  &\Theta\left[ t - t_0 - \biggl|\det_\sigma \biggr| \right]\,.
  \end{split}
\label{Eq:kernel}
\end{equation}
In Eq.~(\ref{Eq:kernel}) $\Theta$ is the Heaviside step function and the spin-dependent $\Delta$-functions are defined as
\begin{subequations}
  \begin{align}
    \det_\sigma (z|z_0) &= \int_{z_0}^{z} \frac{\dd z'}{v_\sigma(z')}\,, \\
    \dett_\sigma (z|z_0) &= \int_{z_0}^{z} \frac{\dd z'}{\tau_\sigma(z')\, v_\sigma(z')}\,.
  \end{align}
\label{Eqs:deltas}
\end{subequations}

Assuming a $\delta$-like source $\Sext_\sigma = S_{\delta\sigma} \, \delta(z-\zs) \delta(t-\ts)$ and $p_\sigma(z) = 0$ we obtain
\begin{equation}
  \begin{split}
    \pder{}{t} n_\sigma(z,t) = &-\frac{n_\sigma(z,t)}{\tau_\sigma(z)} - S_{\delta\sigma}\, \pder{\phi_\sigma(z,t|\zs,\ts)}{z} \\
                              &+ S_{\delta\sigma}\, \delta(z-\zs) \delta(t-\ts)\,.
  \end{split}
\label{Eq:sde2}
\end{equation}
After time discretization we can write the solution at time $t+\delta{t}$ as
\begin{equation}
  \begin{split}
  &n_\sigma(z,t+\delta{t}) =\; e^{-\delta{t}/\tau_\sigma(z)}\, n_\sigma(z,t) \\
		           &-\int_{t}^{t+\delta{t}} \dd t'\; S_{\delta\sigma} \exp\left( - \frac{t' - t - \delta{t}}{\tau_\sigma(z)} \right) \pder{\phi_\sigma(z,t'|\zs,\ts)}{z} \\
		           &+ S_{\delta\sigma} \delta(z-\zs) \delta_{t+\delta{t},\ts}\,.
  \end{split}
\label{Eq:sde3}
\end{equation}

Defining the average particle density over spatial step as
\begin{equation}
  \avn_\sigma(z,t) = \frac{1}{\delta{z}} \int_{z-\delta{z}/2}^{z+\delta{z}/2} n_\sigma(\xi,t) \dd \xi ,
\end{equation}
we can rewrite Eq.~(\ref{Eq:sde3}) as
\begin{equation}
  \begin{split}
  &\avn_\sigma(z,t+\delta{t}) =\; \exp\left(-\delta{t}/\tau_\sigma(z)\right) \avn_\sigma(z,t) \\
  &+ S_{\delta\sigma}\, \left[ \delta_{z,z_0} \delta_{t+\delta{t},t_0} + \psi_{\sigma}^{-}(z,t|z_0,t_0) - \psi_{\sigma}^{+}(z,t|z_0,t_0) \right]\,,
  \end{split}
\label{Eq:sde4}
\end{equation}
where
\begin{equation}
  \begin{split}
    \psi_{\sigma}^{\pm}(z,t|z_0,t_0) = 
    \int_t^{t+\delta{t}} \dd \eta\, &\exp\left( -\frac{\eta - t - \delta{t}}{\tau_\sigma(z)} \right) \\
    &\phi_\sigma \left(z \pm \frac{\delta{z}}{2}, \eta \biggm| z_0,t_0 \right)
  \end{split}
\label{Eq:int_flux}
\end{equation}
are the spin-dependent integrated fluxes.

\subsubsection{General solution}

The general solution for the spin-dependent particle density in a layer at a given energy level $\epsilon$ reads
\begin{equation}
  n_{\sigma}(z,t+\delta{t}) = e^{-\delta{t}/\tau_{\sigma}(z)}\, n_{\sigma}(z,t) + S_{\sigma}^{e}(z,t+\delta{t}) + \Phi_{\sigma}\,,
\label{Eq:gen_sol}
\end{equation}
where the spin-dependent total effective source is
\begin{equation}
  S_{\sigma}^{e}(z, t+\delta{t}) = S_{\sigma}(z, t+\delta{t}) + S_{\sigma}^{p}(z, t+\delta{t})\,,
\label{Eq:gen_source}
\end{equation}
where $S_{\sigma}(z, t+\delta{t})$ is an external contribution to the total source, and
$S_{\sigma}^{p}(z, t+\delta{t})$ is a contribution due to electrons coming from other energy levels $\epsilon'$ with spin $\sigma'$.
The latter term can be calculated as
\begin{equation}
  \begin{split}
    S_{\sigma}^{p}(\epsilon, z, t+\delta{t}) = \sum_{\sigma'} \int_0^{\emax} \dd\epsilon'\; 
    &n_{\sigma'}(\epsilon',z,t)\, p_{\sigma',\sigma}(\epsilon',\epsilon,z,t) \times\\
    &\left( 1 - e^{-\delta{t}/\tau_{\sigma'}(\epsilon',z,t)} \right) \,,
  \end{split}
\label{Eq:Sp_def}
\end{equation}
where $\sigma' \in \{\uparrow,\downarrow\,\leftarrow,\rightarrow\}$, and $p_{\sigma',\sigma}(\epsilon',\epsilon,z,t)$ is the probability that
an electron at energy level $\epsilon'$ and spin $\sigma'$ will move to energy level $\epsilon$ with spin $\sigma$ in the next time step, $t+\delta{t}$.
Moreover $\emax$ is the maximum energy of the hot electrons above the Fermi level that is taken into account.
Finally, the total flux in the spin-channel $\sigma$, $\Phi_{\sigma}$ comprises of right and left-flowing fluxes marked as $\Phi_{\sigma}^{<}$ and $\Phi_{\sigma}^{>}$, respectively.
They are defined as
\begin{subequations}
  \begin{align}
    \Phi_{\sigma}^{<}(z,t) = &\sum_{t_0 = 0}^{t} \sum_{z_0 < z} S_{\sigma}^e(z_0,t_0) \times \notag \\
    &\left[ \psi_{\sigma}^{-}(z,t|z_0,t_0) - \psi_{\sigma}^{+}(z,t|z_0,t_0) \right]\,, \\
    \Phi_{\sigma}^{>}(z,t) = &\sum_{t_0 = 0}^{t} \sum_{z_0 \ge z} S_{\sigma}^e(z_0,t_0) \times \notag \\
    &\left[ \psi_{\sigma}^{-}(z,t|z_0,t_0) - \psi_{\sigma}^{+}(z,t|z_0,t_0) \right]\,,
  \end{align}
\label{Eq:gen_Phi}
\end{subequations}
where, as already mentioned above, the energy-dependence of the quantities is omitted.
The Equations (\ref{Eq:gen_sol}) -- (\ref{Eq:gen_Phi}) 
form the basis for our calculations of the spin-dependent transport through the perpendicular magnetic spin valve
and allow us to calculate the spin current transverse to FM2 magnetization flowing to FM1 as 
\begin{equation}
  \js(z,t) = \frac{\hbar}{2} \int_0^{\emax} {\rm d}\epsilon\, \left[ \Phi_{\leftarrow}^{<}(z,t,\epsilon) - \Phi_{\rightarrow}^{<}(z,t,\epsilon) \right]\,.
\end{equation}
Similarly, replacing $<$ by $>$, one can define current flowing in the opposite direction, $\js'$.

This spin current vanishes in the vicinity of the NM/FM2 interface and the absorbed momentum mostly generates an antidamping spin-transfer torque. \cite{r_02_StilesAnatomy}
The field-like torque due to the STT has been shown to be rather small in transition metals and it is further reduced by any disorder,~\cite{r_07_ct_mixcond} 
hence it will not be considered here. 
The average magnetizations in the first and second layer are labeled as $\bM_1$ and $\bM_2$.
To study the magnetization dynamics we define the current-induced spin torque acting on $\bM_2$ as
\begin{equation}
  \btau_2(t) = -\, \frac{\js(z_{\rm III},t)}{\Ms_2\, V_2}\, \bM_2 \times \left( \bM_2 \times \bM_1 \right)\,,
\label{Eq:stt1}
\end{equation}
where $\js(z_{\rm III})$ is the spin current at the interface number III, and $V_2$ is the volume of the FM2 layer.

The internal interfaces between the layers (II and III) are treated here as reflectionless.
On the other hand, the outermost interfaces (I and IV) completely reflect the electrons which hit these interfaces.


\subsection{Magnetization dynamics}

To model the magnetization dynamics in the FM2 layer 
induced by the spin-transfer torque we make use of the Landau-Lifshitz-Gilbert equation with the spin-torque term, which reads
\begin{equation}
  \begin{split}
    \der{}{\bM_2}{t} = -&\gg \mu_0\, \bM_2 \times {\bm H}_{{\rm eff},2} + \frac{\alpha}{M_{{\rm s} 2}} \bM_2 \times \der{}{\bM_2}{t}\; + \\
    &\frac{\gg}{M_{{\rm s} 2}}\, \btau_2\,,
  \end{split}
\label{Eq:LLG}
\end{equation}
where $\gg = |e| g / (2 m_e)$ is the gyromagnetic ratio with $g=2$ being the Land{\'e} factor, and $e$ and $m_e$ are the electron charge and mass, respectively.
The gyromagnetic ratio is as large as $\gg = 1.7587 \times 10^{11}\, {\rm T}^{-1}{\rm s}^{-1}$.
Moreover,  $\mu_0 = 4\pi \times 10^{-7}\, {\rm N A}^{-2}$ is the vacuum permeability, 
$\alpha$ is the Gilbert damping, 
${\bm H}_{{\rm eff}, 2}$ is the effective magnetic field, and
$\btau$ is the spin torque created by the spin current.
The effective magnetic field is defined as
\begin{equation}
  {\bm H}_{{\rm eff},2} = -\frac{1}{\mu_0 V_{\rm FM2}} \vder{E_2[\bM_2]}{\bM_2}\,,
\label{Eq:Heff}
\end{equation}
where $E_2$ is total energy functional related to the magnetic state of FM2 
and $V_{\rm FM2}$ is volume of the ferromagnetic layer FM2.

The effective magnetic field used in the simulations of the magnetization dynamics is
\begin{equation}
  \begin{split}
    {\bm H}_{{\rm eff},2} = ~ &H_{\rm app}\, \ex + \frac{2\, K_u}{\mu_0\, M_{{\rm s},2}^2 V_{\rm FM2}} \left( \bM_2 \cdot \ex \right) \ex\\
                            &- \frac{2\, K_\perp}{\mu_0\, M_{{\rm s},2}^2 V_{\rm FM2}} \left( \bM_2 \cdot \ez \right) \ez\,,
  \end{split}
\label{Eq:Heff2}
\end{equation}
where the first term stands for the applied in-plane magnetic field with magnitude $H_{\rm app}$, while the second one expresses the magnetic field due to the 
uniaxial easy-axis noncrystalline anisotropy given by the anisotropy constant $K_u$.
The third term introduces the easy-plane anisotropy with constant $K_\perp$.
Vectors $\ex$ and $\ez$ are the unit vectors in the direction of the $x$ and $z$-axes, respectively (see Fig.~\ref{Fig:scheme}).

\section{Results and discussion}
\label{Sec:Results}

\subsection{Simulation methodology}

We study the spin-current flow through the spin valve structure shown in Fig.~\ref{Fig:scheme},
when the laser pulse is applied from the left-hand side to the interface I.
When the laser pulse is applied, the electrons from the $d$-band are uniformly populated in the $sp$-band levels.
In our calculations we assumed $12$ energy levels in the $sp$-band above the Fermi level.
The difference between the energy levels was $\Delta{E} = 0.125\, {\rm eV}$.
The number of excited electrons on the energy level $\epsilon$ with spin $\sigma$ at position $z$, 
$N_\sigma(t,z,\epsilon)$, follows the laser pulse shape.
We assumed a Gaussian-shaped pulse which leads to
\begin{equation}
  N_\sigma(t,z,\epsilon) = \Nbar_\sigma(z,\epsilon)\, \frac{1}{\Delta\, \sqrt{2\pi}}\, \exp \left\{ \frac{(t - t_0)^2}{2\, \Delta^2} \right\}\,,
\label{Eq:Nz}
\end{equation}
where $\Nbar_\sigma(z,\epsilon)$ is the average number of excited electrons at energy level $\epsilon$ with spin $\sigma$ at position $z$.
The position of the pulse peak is given by the time parameter $t_0$ and the width of the pulse is set by parameter $\Delta$.
To make a realistic estimation we assumed a finite penetration depth of the laser pulse.
In practice it means that the average number of excited electrons due to the laser pulse decreases exponentially
as a function of the distance from the interface I with the characteristic length scale given by the laser penetration depth, $\lambda$,
$\Nbar_\sigma(z,\epsilon) = \Nbar_\sigma(0,\epsilon)\, \exp(-z/\lambda)$. 
Setting $\Nbar_{\sigma}(0,\epsilon) = 0.1 \equiv N_0$ for $\sigma = \{\leftarrow, \rightarrow\}$ defines our basic density of excitations. 
The corresponding laser fluence can be estimated as 27.5 mJ\,cm$^{-2}$ for Fe.
In addition, the excitation profile might be modified by multiple reflexions of the laser light from the FM/NM interfaces.
Such an effect might influence the magnitude of the spin currents in the metallic multilayers.
In our calculations, however, we shall focus on understanding of the basic features of the spin torque generation.
Thus, we shall disregard the effects of multiple light reflections.

For the sake of simplicity, we assume both magnetic layers to be of the same material (Fe) while the
central nonmagnetic layer has transport properties typical of Cu. A perpendicular spin valve composed of two thin Fe films has been demonstrated recently.\cite{Razdolski2017:NatComm}
Moreover, we assume the same laser penetration depth for all studied materials of
$\lambda = 15\, {\rm nm}$, which is in good agreement with experimental findings.~\cite{Brown2010:Book,Macdonald:Abstract2003}
In the magnetic layers, electrons are excited just into the spin channels with spin aligned to the magnetization axis.
Thus in FM1, with perpendicular magnetization, spin are excited in the spin channels $\leftarrow$ and $\rightarrow$ while
for FM2 the laser pulse induced electrons appear in the channels $\uparrow$ and $\downarrow$.
On the other hand, in the nonmagnetic layer, spins are homogeneously populated into all four spin channels.
The average number of excited electrons at position $z$ is initially assumed to be uniform in the energies,
$\Nbar_\sigma(z,\epsilon_i) = \Nbar_\sigma(z,\epsilon_j)$ for all $i, j = 1, 2, \dots, 12$.
Moreover, the average number of excited electrons are spin symmetric, which leads to
$\Nbar_{\leftarrow}(z,\epsilon) = \Nbar_{\rightarrow}(z,\epsilon)$ with $\Nbar_{\uparrow}(z,\epsilon) = \Nbar_{\downarrow}(z,\epsilon) = 0$ in FM1, 
$\Nbar_{\uparrow}(z,\epsilon) = \Nbar_{\downarrow}(z,\epsilon)$ with $\Nbar_{\leftarrow}(z,\epsilon) = \Nbar_{\rightarrow}(z,\epsilon) = 0$ in FM2, and
$\Nbar_{\leftarrow}(z,\epsilon) = \Nbar_{\rightarrow}(z,\epsilon) = \Nbar_{\uparrow}(z,\epsilon) = \Nbar_{\downarrow}(z,\epsilon)$ in NM.
Consequently, the excited electrons move according to the transport equations introduced in Sec.~\ref{Sec:Model}.
Although the number of excited electrons in the magnetic layers is equal in both spin channels, 
the spin current builds up due to different electron velocities and relaxation times in the two longitudinal spin channels of a ferromagnetic layer.

If we replace the energy dependence of electrons, crucial for the ultrafast aspects of the problem, by the Fermi-Dirac distribution, 
our model would become equivalent to a composition of nonmagnetic part containing majority and minority electrons with spin polarization axis 
either along the magnetization of the FM1 or FM2 layer, 
and a magnetic part where only spins aligned with the local magnetization effectively contribute to transport, 
as already presented by Slonczewski~\cite{r_02_Slon_CurrTor}.

\subsection{Ultrafast demagnetization}

\begin{figure}[tp!]
  \includegraphics[width=.9\columnwidth]{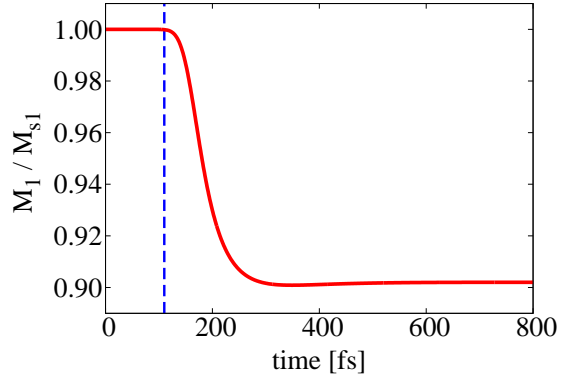}
  \caption{(Color online) Calculated ultrafast demagnetization of the FM1 layer in a spin valve FM1(6)/NM(2)/FM2(6) after applying a laser pulse 
           with fluence corresponding to $\Nbar_{\sigma}(0,\epsilon) = 2 N_0$ at interface I.
           The position of the peak is shown by the blue dashed line. The pulse length was as large as $\Delta = 40\, {\rm fs}$.
           $M_{\rm s1}$ is the magnetization saturation value of FM1 in equilibrium.
           In the calculations FM1 and FM2 was assumed to be Fe, while NM corresponds to Cu.}
\label{Fig:demag}
\end{figure}
After applying the laser pulse with $\Delta = 40\, {\rm fs}$ to the interface I, 
one observes an ultrafast demagnetization in FM1 layer. 
For a spin valve FM1(6)/NM(2)/FM2(6), where the numbers in the brackets are the layer's thicknesses in nanometers, 
we assume the number of excited electrons at interface I to be $N_0$.
The computed reduction of the magnetization in FM1 is shown in Fig.~\ref{Fig:demag}.
We observe an average demagnetization of FM1 as large as $10\%$.
As a result of FM1s demagnetization, electrons in channels featuring spins aligned with the FM1 magnetization ($\leftarrow$ and $\rightarrow$)
start to move through the heterostructure. During the transport electrons scatter and thereby relax towards lower energy levels,
which causes a continuous decay of the spin current ($\js = j_{\leftarrow} - j_{\rightarrow}$).
In the rest of this section we shall focus on this spin current calculated at the NM/FM2 interface
since this is the direct measure of the spin torque acting on the FM2 magnetization.

\subsection{Superdiffusive spin-transfer torque}

\begin{figure}[tp!]
  \includegraphics[width=.9\columnwidth]{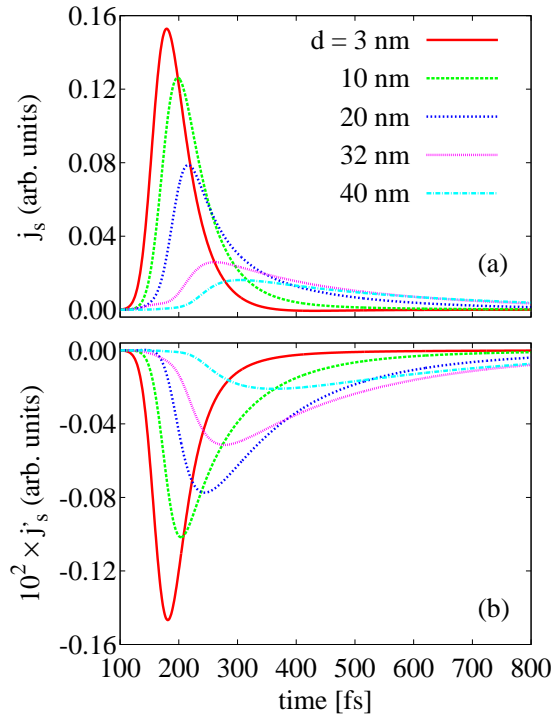}
  \caption{(Color online) Calculated superdiffusive spin currents transverse to the FM2 magnetization at the NM/FM2 interface (III) in FM1(6)/NM($d$)/FM2(6) spin valves    
           (a) flowing from left to right,
           (b) flowing from right to left.
           A laser fluence corresponding to $\Nbar_{\sigma}(0,\epsilon) = 2 N_0$ is assumed.}
\label{Fig:NM_thick_time}
\end{figure}

First, we study the temporal dependence of the transverse spin current at the NM/FM2 interface.
Figure ~\ref{Fig:NM_thick_time}(a) depicts the time evolution of the superdiffusive transverse spin current flowing from left to right at the interface III
calculated for the FM1(6)/NM($d$)/FM2(6) structure for various thicknesses of the NM layer.
As a consequence of the Gaussian-shaped laser pulse, the typical time evolution of the spin current has a peak, which
decreases with the thickness of the NM layer. In addition, the peak position also shifts in time, 
the peak width increases, and the tails become longer.
These features are a result of the longer distance which electrons have to pass through until they meet the NM/FM2 interface.
On the other hand, a nonzero transverse spin current at the NM/FM2 interface can be also observed in the direction from right to left.
This spin flow is caused by the avalanches of electrons which are triggered by collisions of hot electrons with electrons in localized atomic orbitals.
Fig.~\ref{Fig:NM_thick_time}(b) shows the time dependence of the opposite current at the NM/FM2 interface.
The time-dependence of the opposite current has a similar shape as the direct one.
Importantly, the magnitude of the opposite spin flow is about two orders smaller than the one moving towards the interface III.
In order to compare the spin torque acting on FM2s magnetization in different multilayers, 
we calculate the total transverse spin momentum transferred across the NM/FM2 interface, given as
\begin{equation}
  \Delta S = \int_{0}^{t_{\rm max}} j_{\rm s}(z_{\rm III},t)\; {\rm d}t\,,
\end{equation}
where $t_{\rm max}$ was taken as large as $2\, {\rm ps}$.
Analogously, one can define the total spin current, $\Delta S'$, flowing in the opposite direction.

\begin{figure}[tp!]
  \includegraphics[width=.9\columnwidth]{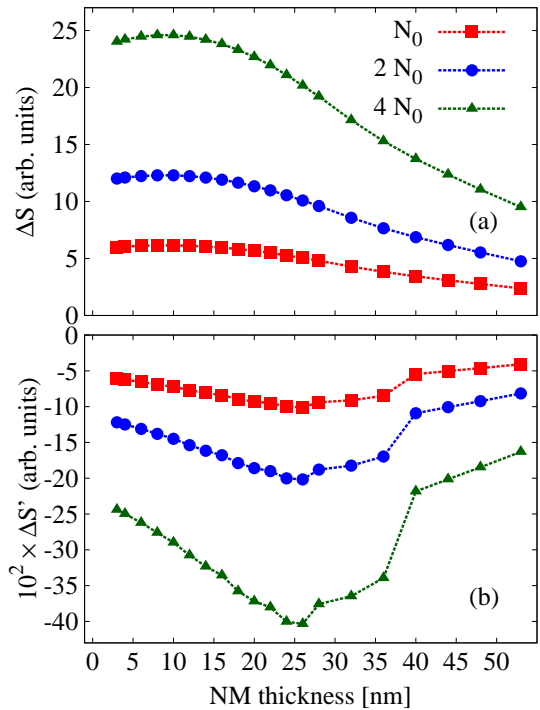}
  \caption{(Color online) Total transverse spin momentum transferred across the NM/FM2 interface (III) transverse to the FM2 magnetization as a function of the NM thickness 
           (a) flowing from left to right,
           (b) flowing from right to left.
           The thickness dependence is calculated for three different laser fluences corresponding to $\Nbar_{\sigma}(0,\epsilon) = N_0$, $2N_0$, and $3N_0$.}
\label{Fig:NM_thick}
\end{figure}

Fig.~\ref{Fig:NM_thick} shows the total transverse spin momentum at the NM/FM2 interface
in the direction from left to right (a), and from right to left (b), calculated for
various thicknesses of the NM layer.
For smaller NM thicknesses, the transverse total spin current flowing from the left is almost constant.
However, a slightly nonmonotonous thickness dependence can be observed. 
Namely, there is a maximum total transverse spin current which appears at $d \sim 10\, {\rm nm}$.
For larger NM thickness the integrated spin current decreases.
The reason of this nonmonotonous thickness dependence is both the relatively long laser penetration depth
as well as electron reflexions from the outermost interfaces (marked as I and IV in Fig.~1).
In more detail, electrons excited by the laser move in both directions.
Thus, even a part of the charge current excited in the nonmagnetic layer can polarize in the FM1 layer and
after reflexion from the left interface (I), it can contribute to the spin torque acting on
FM2s magnetization.

\subsection{Generation of the spin current}

The most important aspect for laser-operated spin devices is the generation of the SC. 
In the here-studied geometry the main source of the nonthermal spin current is the laser-excited FM1 layer. 
Some experiments suggest different ways  as to how the superdiffusive spin current generation is distributed along the multilayer, especial along FM1.
Alekhin {\em et al.}~\cite{Alekhin2015} deduce that in a layer consisting of Fe/Au films just a very thin interfacial region (around 1 nm) contributes to the spin current generation.
On the other hand, Lalieu {\em et al.}~\cite{Lalieu2017:PRB} show that, at least in a small range of thicknesses, 
almost the whole out-of-plane magnetized layer,
consisting of Co/Ni thin films, is used for SC generation. 
In this limit, one should expect a different behavior of the spin current acting on FM2
when the thickness of FM1 is increasing. 
These two cases establish the boundaries of the dynamics presented here,
where the density of excited hot electrons decays exponentially from the laser spot as given by Eq.~(\ref{Eq:Nz}).

\begin{figure}
  \includegraphics[width=.9\columnwidth]{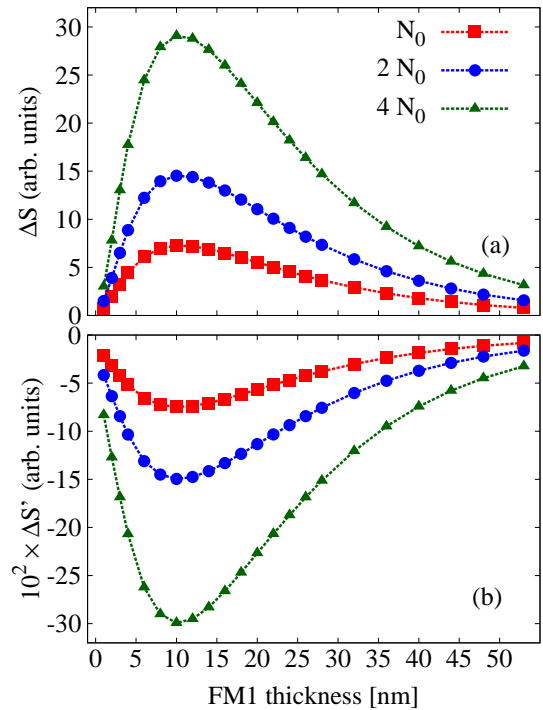}
  \caption{(Color online) Total transverse spin momentum transferred across the NM/FM2 interface (III) transverse to the FM2 magnetization as a function of FM1 thickness 
           (a) flowing from left to right,
           (b) flowing from right to left.
           The thickness dependence is calculated for three different laser fluences corresponding to $\Nbar_{\sigma}(0,\epsilon) = N_0$, $2N_0$, and $3N_0$.}
\label{Fig:FM1_thick}
\end{figure}
One of the possible expectations for the superdiffusive model is that
when the thickness of FM1 is increasing, the current traveling through the FM1 layer will encounter stronger spin filtering,
which would result in a higher spin current leaving the polarizer.
However, there are a few factors working against this assumption.
If the probability of the spin-flip processes is small, the spin current in FM1 appears
due to different electron velocities and lifetimes in the two spin channels.
Moreover, during the transport the laser-excited electrons encounter relaxation processes (i.e., scattering) that lower their energies, which are stronger in the magnetic layers than in the NM.  
As a result, the particle and the spin currents approaching the NM/FM2 interface will decrease with FM1 thickness.
This would limit the effective thickness of FM1 to the electron relaxation lengthscale.
A second factor is the finite laser penetration depth, which also limits the efficiency of the spin-current source.
Our results calculated for a spin valve FM1($d$)/NM(6)/FM2(6) are shown in Fig.~\ref{Fig:FM1_thick}.
The plotted results suggest that the thickness of FM1 layer for an optimal spin-torque efficiency is
about $\sim 10\, {\rm nm}$. When the FM1 thickness goes beyond this value, the transverse spin current
at the NM/FM2 decreases. 
A similar behavior is obtained for the current in opposite direction, 
which, however, is about two orders of magnitude smaller than the direct one.
This optimum thickness of the FM1 layer depends on the laser penetration depth as well as the FM1 material parameters.
However, it is independent on the thickness of NM layer. 
Overall, are findings are qualitatively in agreement with the measurements of Lalieu {\em et al.},~\cite{Lalieu2017:PRB} who have observed an increase of the canting angle in the FM2 layer with  FM1 thickness within the examined thickness range below 5 nm (i.e., up to 4 Co/Ni repetitions).

\subsection{Magnetization dynamics}

Having calculated the magnitudes of the superdiffusive spin torque, we can examine its effect on the magnetization dynamics in the studied spin valve.
Since we calculated the electron transport for the static magnetic configuration, we expect just minor spin dynamics
in the FM2 layer.
In the above calculations we have shown that the transverse spin current at the FM1/NM interface flowing from right to left
is much smaller than the one at the NM/FM2 layer flowing from left to right.
Therefore, we assume that the dynamics of the FM1 magnetization is negligible in comparison to the FM2 one.

To study the magnetization dynamics we assume a homogeneous magnetization in the FM2 layer, 
which can be described by a macrospin model.
We note 
that this assumption does not provide access to all the experimentally observed
details like e.g.\ THz spin waves observed close to the NM/FM2 interface,~\cite{Lalieu2017:PRB,Razdolski2017:NatComm} whose simulation would go beyond the scope of this article. For small FM2 thicknesses spin waves are not observed \cite{Lalieu2017:PRB} and the error made by averaging over the FM2 layer becomes negligible.
The dynamics of the macrospin is described by the LLG equation with in-plane (Slonczewski) term~\cite{Slonczewski1996}
with a time-dependent current density.
In the effective magnetic field we assume an
in-plane uniaxial magnetic anisotropy with equilibrium energy $K_{\rm u} = 0.05\, {\rm mRy}$ per atom,
$K_\perp \simeq 0.04\, K_{\rm u}$ and a static external magnetic field applied in the layer's plane
along the equilibrium position of the magnetization as large as $\mu_0\, H_{\rm app} = 100\, {\rm mT}$
(see Eq.~(\ref{Eq:Heff2})).
Moreover, we have assumed a Gilbert damping parameter as large as $\alpha = 0.1$.

\begin{figure}[t!]
  \includegraphics[width=.9\columnwidth]{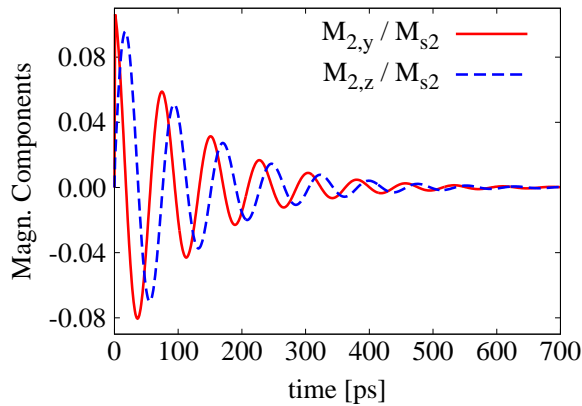}
  \caption{(Color online) Laser-induced magnetization dynamics in the FM2 layer calculated for the spin valve FM1(16)/NM(4)/FM2(6).
           The solid (red) and dashed (blue) lines are the two transverse components of the magnetization. A laser fluence corresponding to $\Nbar_{\sigma}(0,\epsilon) = N_0$ is assumed.
          }
\label{Fig:dynam}
\end{figure}
To validate the feasibility of the theoretical framework presented here, 
we computationally reproduce the experimental heterostructure studied by Razdolski \textit{et al.}~\cite{Razdolski2017:NatComm}  
There, a noncollinear Fe/Au/Fe spin valve with a FM1 thickness of 16 nm is pumped by a laser with fluence 10 mJ\,cm$^{-2}$, 
producing an initial change of the FM2 magnetization due to the superdiffusive STT equal to $\Delta M_{2}=0.02\, {M_{{\rm s} 2}}$. 
The results of our simulations, performed adjusting the excitation density to the experimental fluence, 
provide  $\Delta M_{2}=0.034\, {M_{{\rm s} 2}}$, in good agreement with the experimental result. 
It is worth noticing that this value depends also on the thickness of the FM2 layer where the magnetization change is assumed to distribute.

The results of our LLG simulations of the FM2 magnetization dynamics for a sample spin valve FM1(16)/NM(4)/FM2(6) 
are plotted in Fig.~\ref{Fig:dynam}.
The laser pulse is calculated to excite a homogeneous magnetization dynamics in FM2 with frequency $\approx 10\, {\rm GHz}$.
These precessions are exponentially damped and vanish after $\approx 600\, {\rm ps}$,
similarly to what has been observed in the recent experiments.~\cite{Lalieu2017:PRB}

\section{Conclusions}
\label{Sec:Conclusions}

We have presented a theoretical study of the laser-induced spin-transfer torque and concomitant 
magnetization dynamics in a spin valve composed of two magnetic layers with noncollinear magnetizations
separated by a normal metal.
To this end we have extended the model of superdiffusive transport~\cite{Battiato2010:PRL}
to account for noncollinear magnetic configurations.
In the bulk of the magnetic layers we separate the nonthermal spin current with respect to the local magnetization direction into a longitudinal and a transverse part.
While the longitudinal spin current pulse is generated and continues into the second ferromagnet, the transverse component is rapidly absorbed in FM2 giving rise to the spin-transfer torque.
The proposed simplified model is restricted to homogeneous magnetizations inside the magnetic layers and
relatively slow magnetization dynamics limited to small angle precessions around the equilibrium.
Moreover, in our study we have assumed only fully transparent interfaces between the layers without any reflections. We note, however, that specific reflection and transmission coefficients could be incorporated in the superdiffusive transport model.

We have applied the model to study computationally the spin-dependent transport in the FM1/NM/FM2 spin-valve structure.
Particularly, we have focused on the spin-transfer torque (spin current) acting on the FM2 magnetization
as a function of the spin-valve geometry.
An exponential decrease of laser-generated electrons with the depth has been included in our calculations. 
Importantly, we have shown that it leads to the existence of an optimal thickness of the FM1 for which the total (time-integrated) spin torque acting on the FM2 magnetization is maximal.
For a combination of Fe and Cu layers we identify this thickness to be $10\, {\rm nm}$.
When the thickness of the FM1 layer exceeds the optimum, the spin current at the NM/FM2 interface decreases.

Finally, we have used the calculated laser-induced spin-transfer torque to simulate the magnetization dynamics in the
second magnetic layer employing the macrospin approximation.
Using the Landau-Lifshitz-Gilbert equation we have shown that the spin-torque action occurring on a sub-ps timescale is sufficient to trigger small angle magnetization precessions, which can persist for few hundred picoseconds.
This result is consistent with recent experimental observations.~\cite{Schellekens2014:NatComm,Choi2014:NatComm,Lalieu2017:PRB}

The here-developed noncollinear superdiffusive transport theory forms a basis for future numerical investigations 
of ultrafast spin-transfer torques and spin torque induced magnetization dynamics in metallic heterostructures.

\section*{Acknowledgement}

We acknowledge financial support from the Czech Science Foundation (Grant No.\ 15-08740Y), the Swedish Research Council (VR),  
the K.\ and A.\ Wallenberg Foundation (Grant No.\ 2015.0060), and the European Union's Horizon2020 Research and Innovation Programme  
(Grant agreement No.\ 737709,  FEMTOTERABYTE). We thank M.\ Battiato for useful discussions.
Access to computing and storage facilities owned by parties and projects contributing to the National Grid Infrastructure MetaCentrum 
provided under the programme ``Projects of Large Research, Development, and Innovations Infrastructures" (CESNET LM2015042), is greatly appreciated,
as well as support from the Swedish National Infrastructure for Computing (SNIC).


%

\end{document}